\documentclass[twocolumn,trackchanges]{aastex631}

\usepackage{amssymb}

\begin{document}

\title{Wakes from Companion Interactions in Type Ia Supernovae Nebular Emission Line Profiles}

\author[0009-0000-1833-1746]{Kathlynn Simotas}
\affiliation{Department of Physics, University of California, Santa Barbara, CA 93106, USA
}

\author[0000-0001-8038-6836]{Lars Bildsten}
\affiliation{Department of Physics, University of California, Santa Barbara, CA 93106, USA
}
\affiliation{Kavli Institute for Theoretical Physics, University of California, Santa Barbara, CA 93106, USA
}

\author[0000-0002-7174-8273]{Logan J. Prust}
\affiliation{Kavli Institute for Theoretical Physics, University of California, Santa Barbara, CA 93106, USA
}

\begin{abstract}

 Thermonuclear supernovae (SNe) are the result of the nuclear transformation of carbon/oxygen (C/O) white dwarfs (WDs) to the radioactive element $^{56}\mathrm{Ni}$ and intermediate mass elements (IMEs) like Ca, Ar, etc. Most progenitor scenarios involve a companion star which donates matter to the exploding white dwarf, implying a fundamental prediction: the formation of a wake in the explosive ejecta as it runs into and moves past the companion star. This wake leaves an indelible imprint on the ejecta's density, velocity, and composition structure that remains fixed as the ejecta reaches homologous expansion. We simulate the interaction of the ejecta and Roche-lobe filling donor in a double degenerate double detonation Type Ia progenitor scenario and explore the detectability of this imprint in late-time nebular phase spectroscopy of Type Ia SNe under the assumption of local heating  ($t > 200$ days). At these times, the velocity profiles of forbidden emission lines reflect the velocity distribution of all of the ejecta and the critical electron density for that forbidden line. We explicitly calculate line shapes for the [Co III] $11.89 \mu\mathrm{m}$ line that traces the initial $^{56}\mathrm{Ni}$ distribution and the [Ar III] $8.99 \mu\mathrm{m}$ line, which traces a typical intermediate mass element. We predict the viewing angle dependence of the line shape, present a tool to quickly calculate optically thin line shapes for various 3D density-velocity profiles and discuss JWST observations.

\end{abstract}


\section{Introduction} \label{sec:intro}

Despite their importance as cosmic distance indicators and probes of stellar evolution, the progenitor systems of Type Ia supernovae (SNe Ia) are still the subject of intense debate. It is commonly accepted that SNe Ia are caused by carbon-oxygen white dwarfs (CO WDs) accreting mass from a  companion donor star, resulting in a thermonuclear explosion. However, various progenitor models and explosion models still struggle to account for observed variations in SNe Ia characteristics \citep{liu2023type}. Nearly all progenitor scenarios feature accretion onto a CO WD, implying an interaction between the exploding white dwarf’s ejecta and its companion. This is inevitable regardless of the nature of the companion or its survival of the explosion. Quantifying the nature of these interactions may offer a promising complement to existing work discriminating among progenitor channels \citep{marietta2000type,tanikawa2019double}. 

Spectroscopy of the nebular phase of ejecta (hundreds of days after explosion, once the ejecta has become largely optically thin) provides a uniquely clean window for this exploration. Although nebular modeling requires difficult and computationally expensive non-LTE calculations \citep{shen2021non}, late time spectra allow direct and detailed studies of the composition and densities of ejecta which are significantly impacted by companion interactions. Historically, progress in nebular phase spectroscopy has been limited by a paucity of late-time mid-infrared spectra (only seven such data sets existed prior to the debut of the James Webb Space Telescope) \citep{ashall2024jwst, gerardy2007signatures}. JWST has triggered a renaissance in nebular-phase studies, delivering orders-of-magnitude improvements in spectral coverage and sensitivity in the previously underexplored mid-IR, providing a probe of the innermost regions of SNe Ia ejecta \citep{derkacy2023jwst,ashall2024jwst,kwok2023jwst,kwok2024ground,kwok2025jwst}.

In this work we develop a baseline physical understanding of ejecta–companion interaction and its impact of nebular-phase spectra. We quantify the velocity and density perturbations imprinted by companion interaction, develop a code to predict their observable signatures in the mid-IR nebular spectrum, and discuss implications and future extensions. By confronting these predictions with forthcoming JWST observations, we aim to place new constraints on the elusive progenitor problem of SNe Ia. 

Modeling companion interaction within a Roche-lobe filling binary system is applicable across various sub-Chandrasekhar progenitor scenarios. In the dynamically-driven double-degenerate double-detonation (D6) scenario \citep{shen2018three}, two white dwarfs undergo unstable mass transfer and a thin layer of helium on the primary detonates, triggering a second detonation in the core of the accretor. The surviving donor is ejected at $\approx 1,000 \ \mathrm{km/s}$, naturally explaining the recently discovered population \citep{shen2018three}. Recent surveys of hypervelocity white dwarfs moving at speeds comparable to those predicted provide strong evidence that at least some SNe Ia are formed this way and reinvigorated interest in the D6 scenario \citep{shen2018three, el2023fastest}. WDs undergoing stable helium accretion can also undergo a double detonation and lead to remnant stars moving at comparable velocities \citep{wong2023dynamical}.

The `thick' helium-shell model, where a CO WD accretes a larger mass of He from a helium star companion, is another clear application of companion interaction since the helium star companion is always predicted to survive the explosion of the white dwarf \citep{bauer2019remnants}. Due to the large amount of ash from the He shell, this scenario leads naturally to the `Ca-strong' (often `Ca-rich') nebular spectra predicted for low-mass explosions \citep{bauer2019remnants, polin2021nebular, de2019ztf,de2020zwicky}. 

These scenarios demonstrate the importance of developing  quantitative predictions of the impact of ejecta interaction with Roche-lobe-filling companions. Previous works \citep{prust2025ejecta} highlighted the effect on the supernovae remnant whereas this paper focuses on the late nebular emission line profiles. In Section \ref{sec:interact}, we discuss the Athena++  simulation, the collision dynamics and how the companion interaction permanently modifies the ejecta. Section \ref{sec:emission}  models the non-LTE nebular emission line shapes, and tests our work  against JWST data and the 1D non-LTE radiative transfer  code of \cite{blondin2023nebular}. Section \ref{sec:wakes} uses our new code alongside our simulated ejecta with wakes to calculate line shapes for representative lines in the iron group and IMEs.  We discuss and conclude in Section \ref{sec:conclusion}. \ref{sec:A1} highlights some nuances in our discussion of Section \ref{sec:wakes}.

\section{Companion Interaction Simulation } \label{sec:interact}

\subsection{Simulation Setup }

We perform hydrodynamical simulations using Athena++ \citep{Stone2020} of the ejecta-companion interaction for a double-degenerate scenario where the donor survives. We use a 3D spherical-polar static mesh in the same way as \cite{prust2025ejecta}, placing the explosion at the origin and a donor of radius $R = 0.0795R_{\odot}$ at $\theta = 0$ and at the orbital separation $a = 0.269R_{\odot}$. The simulation domain extends from an inner radius at $R_{in} = 0.129R_{\odot}$ where the ejecta is initial injected, out to a radius of $R_{out} = 10.7R_{\odot}$. The domain extends to a maximum polar angle of $80.2^\circ$ and an azimuthal angle of $360^\circ$. The domain includes $N_r \times N_\theta \times N_\phi = 2160 \times 200 \times 20 = 8{,}640{,}000$ cells. We place ambient material throughout the domain at a density of $\rho_0 = 10^{-6}\,\mathrm{g\,cm}^{-3}$
 and pressure $P_0 = 5 \times 10^6\,\mathrm{dyne\,cm^{-2}} 
$, allowing the ejecta to expand into a medium with a mass more than 3 orders of magnitude lower than the total ejecta mass. More details on the simulation setup and nature of the Athena++ solvers can be found in \cite{prust2025ejecta}. 

We inject at the inner boundary of the domain the  Gaussian ejecta model of \cite{wong2024shocking}: 
\begin{equation}
    \rho_G(v, t) = \frac{M_{\mathrm{ej}}}{(v_0 t \sqrt{\pi})^3} \exp\left(-\frac{v^2}{v_0^2}\right),
\end{equation}
where the characteristic velocity is
\begin{equation}
    v_0 = \sqrt{\frac{4E_{\mathrm{ej}}}{3M_{\mathrm{ej}}}}.
\end{equation}
The Gaussian model is parameterized by $E_{ej}$, the total energy of the ejecta, and $M_{ej}$ the total ejecta mass which we set to $E_{ej} = 0.97 \times 10^{51} \mathrm{ergs}$ and $M_{\mathrm{ej}} = 0.9 M_{\odot}$  to match \cite{prust2025ejecta} and earlier work \citep{bauer2019remnants, wong2024shocking}. We find, as we will discuss in Sections \ref{subsec:34} and \ref{sec:conclusion}, that the Gaussian model also provides an excellent prediction for the undisturbed nebular line profiles. 

Following \cite{prust2025ejecta}, we employ the same equation of state and closure relation that accounts for both gas and radiation pressure as the ejecta evolves, treating the gas as ideal and assuming local thermodynamic equilibrium for the radiation. We also set the initial pressure and model our donor star as a `hard target' or reflective spherical boundary also using the same methods as in \cite{prust2025ejecta}. As the authors mention in that work, this treatment of the donor is physically motivated particularly at high velocities and large radii because tidal deformation and stripping of the donor were found to have minimal impact on the overall ejecta of the double detonation scenario in \cite{wong2024shocking}.

\begin{figure}[ht!]
\centering
\includegraphics[
  width=0.49\textwidth,
  trim= 35 0 0 0,
  clip
]{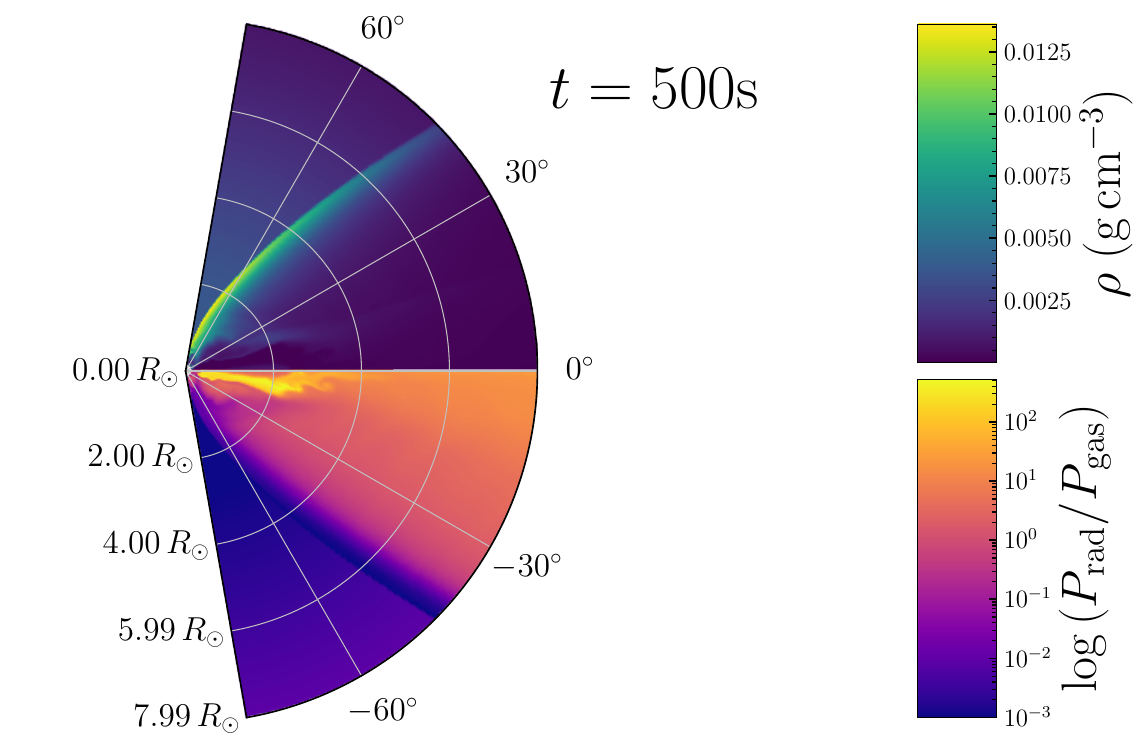}
\caption{The ejecta properties of the simulation at $\phi = 0$ which is representative of the average ejecta properties. The top color bar and half of the polar plot display the density at $500\mathrm{s}$ post-explosion. The bottom color bar and lower half of the polar plot shows the log of the ratio of radiation and gas pressure. The low density wake is present in both density and pressure space, and the bow shock is clearly visible as an overdensity and the border between gas and radiation pressure dominated regions of the ejecta. 
\label{fig:ejecta_properties}}
\end{figure}

\subsection{Consequences of Companion Interaction}

The collision between the ejecta and donor creates a bow shock which deflects the fluid away from the axis of symmetry, creating a low-density, shock-heated wake. The resulting shock cone extends to $\pm 30^\circ$ from the symmetry axis, or $10\%$ of $4\pi$ steradians. A portion of the shocked ejecta also passes through a recompression shock downstream of the donor, further altering its thermodynamic properties. 
As shown in Figure \ref{fig:ejecta_properties}, by 500s post-explosion, the singly-shocked ejecta within $\approx 40^\circ$ of the explosion axis is radiation pressure dominated, while the un-shocked ejecta is gas pressured dominated. The doubly-shocked material that comprises the low-density wake is the most extremely radiation pressure dominated. 
Previous work has either directly discussed the existence of a low-density wake \citep{kasen2004could}, or indirectly alluded to it in figures \citep{tanikawa2019double,pollin2024fate, boos2024type}  but little exploration of the lasting consequences of the wake have been made.

\begin{figure}[ht!]           
  \includegraphics[width=0.49\textwidth,
  trim= 35 0 0 0,
  clip]{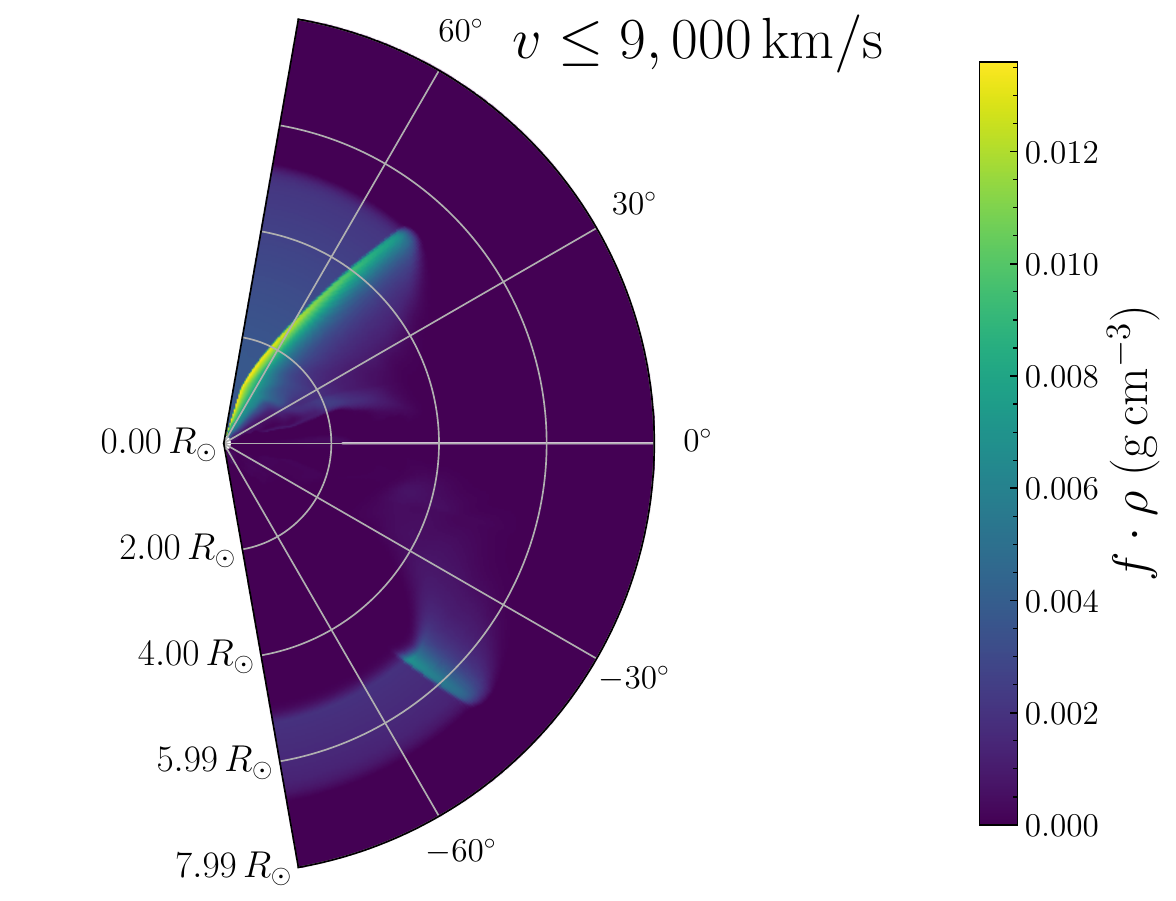}
  \caption{The first two velocity-binned passive scalar quantities (in fraction from 0 to 1 with 0 being none of that bin material present and 1 being ejecta entirely made of that bin's material) multiplied by the mass density. The top shows the lowest velocity bin which is less than $7,000 \mathrm{km/s}$ and the innermost region of the ejecta. The bottom shows the second (and first shell-like) velocity bin from $7,000 - 9,000 \mathrm{km/s}$. The scalar or tracer fractions are multiplied by density to emphasize the amount of  material present. \label{fig:tracer_density}}
\end{figure}

To explore the impact of the wake on composition, we introduce five passive scalars (scalar quantities evolved in a Lagrangian sense)  to our simulation as proxies for tracking the ejecta composition. We bin our ejecta at the injection point into five radial velocity bins motivated by the double detonation compositions in velocity space from the top panel of Figure 7 of \cite{boos2021multidimensional}. Our lowest bin includes material initially moving at $v \leq 7,000 \mathrm{km/s}$. The next bin is $7,000 < v \leq 9,000 \mathrm{km/s}$, then $9,000 < v \leq 11,000 \mathrm{km/s}$, $11,000 < v \leq 14,000 \mathrm{km/s}$ and finally our highest bin is $v  > 14,000 \mathrm{km/s}$. We use these velocity-binned passive scalars, from here on known as `velocity tracers', to follow ejecta material as the ejecta evolves. These tracers, which range from 0 to 1, are local mass fractions for the aforementioned velocity bins and determine the extent to which material is mixed between regions of the ejecta and give a velocity space analog to a composition profile.

Figure \ref{fig:tracer_density} shows the two lowest velocity bin tracers (in their fractional form $f$) multiplied by the mass density of the ejecta. This demonstrates the full distribution of the heaviest two groups of elements throughout the ejecta at 500s post explosion, well into the homologous regime. There is a distinctive bow shock present in the tracer densities that is characteristic of the wake. This bow shock is still present in higher velocity shells of material such as the bottom half of Figure \ref{fig:tracer_density} as well as an overdensity extending radially throughout the shell at about $40^\circ$. Figure \ref{fig:tracer_density} also demonstrates that the ejecta within the wake is at least a factor of a few less dense than in the large angle undisturbed ejecta and an order of magnitude less dense than along the bow shock itself. This density structure is the indelible imprint of the companion interaction and will remain present as the SNe is observed as well as substantially impact the nebular line shapes.

\begin{figure*}[ht]           
  \centering
  \includegraphics[width=\textwidth]{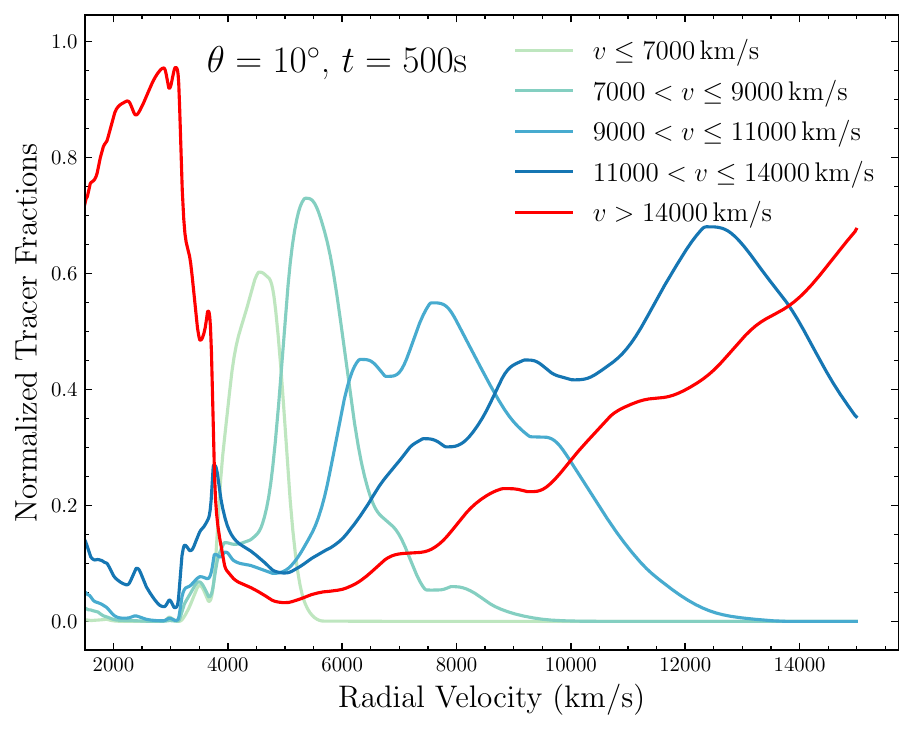}
  \caption{Composition profiles of passive scalars  in velocity space at an angle of $10^\circ$ relative to the explosion axis. The red line indicates the element group that was initially moving at the highest velocities, which is now highest in the innermost region moving at $<4,000\mathrm{km/s}$. All other groups remain in the order if not the exact velocity range of their initial distribution. \label{fig:composition}}
\end{figure*}

 We also find that the slowest moving material in the low density wake is the originally fastest moving ejecta that was trapped behind the donor. Figure \ref{fig:composition} shows this surprising consequence, an inversion of the ordering of our velocity tracers in velocity space for small angles in the wake. There is a clear discrepancy characteristic of the companion interaction: the presence of a significant proportion of lighter elements (which were originally in the highest velocity bin) in the lowest velocity region of the wake. This composition inversion is also  detectable in the tracer density plots in Figure \ref{fig:tracer_density} as a lack of slow moving material in the wake at less than $20^\circ$. The presence of this light material, coupled with the much larger ratio of $P_{rad}/P_{gas}$ in the bottom of Figure \ref{fig:ejecta_properties} point to a cohesive explanation of the innermost low density wake as doubly-shocked material from just after the explosion  that immediately fills the region 
 behind the donor. Finding this slow-moving material is likely a challenge, as it is substantially lower in density (and hence mass), and may be mixed with material lost by the donor at comparable velocities 
\citep{wong2024shocking}.


\section{Nebular Emission Line Profiles} \label{sec:emission}

The nebular phase of SNe is characterized by the free escape of optical and near-infrared photons due to the low opacity of the expanded material. Line formation initially occurs for permitted resonance lines, with forbidden lines emerging later as the Sobolev optical depth decreases sufficiently \citep{sobolev1957diffusion,Jerkstrand_2017}. After a hundred days most optical and near-IR lines have entered the nebular phase. Around the same time, the assumption of local thermodynamic equilibrium (LTE) breaks down due to low densities as $\rho \propto  t^{-3}  $, and many of the simplifying assumptions common in photospheric-phase radiative transfer no longer apply. While this complicates modeling, the nLTE conditions also enable the emergence of forbidden lines that are otherwise suppressed in denser environments \citep{Jerkstrand_2017}. Due to the relative difficulty, rarity, and computational expense of full 3D radiative transfer codes, the state of the art in non-LTE radiative transfer for supernovae spectra are 1D non-LTE codes assuming steady state as exemplified in \cite{blondin2023nebular}.

\subsection{Modeling Nebular Emission Lines} \label{subsec:31}

During the nebular phase, heating via the radioactive decay of ${}^{56}\mathrm{Co} \rightarrow {}^{56}\mathrm{Fe}$ dominates as the initial ${}^{56}\mathrm{Ni}$ has already decayed. Early in this phase ($\leq200$ days), energy deposition is mostly from gamma rays produced in the decay of ${}^{56}\mathrm{Co}$.  However, as the ejecta expands, its optical depth to gamma rays drops rapidly, leading to an increasing fraction of gamma rays escaping without depositing energy \citep{milne1999positron}, so that positrons emitted in ${}^{56}\mathrm{Co}$ $\beta$ decays become the primary source of local energy deposition \citep{milne1999positron,desai2025plasma}. This transition has significant implications for the ionization state and free electron density within the ejecta.  

To estimate the emissivity of our own simulation we rely on the models of \cite{blondin2023nebular}, calculated for a double detonation (D6) model (the most similar to our own with a surviving companion). \cite{blondin2023nebular} used 
 the  spherically averaged density profiles from 3D simulation first introduced in \cite{gronow2021double}. Referred to as `$\texttt{M10\_02}$' in \cite{gronow2021double} or `DBLEDET' in \cite{blondin2023nebular}, this model originated with a 1.025 $M_{\odot}$ total mass, with a 1.005 $M_{\odot}$ core mass. We used an explosion energy $E_{exp} \approx 1 \times 10^{51} {\rm ergs}$. 
 
 The green lines in Figure 3 of \cite{blondin2023nebular} display the ejecta properties: gas temperature $T$, electron density $n_e$, and mean ionization states at 270 days post explosion. They found that the ejecta is nearly isothermal across the full range of velocities and that the ionization state was nearly uniform for many elements 
  \citep{blondin2023nebular}, implying 2 free electrons per atom. 
  As the line emissivity depends primarily on $T$, ionization state, and $n_e$, their demonstrably isothermal profiles and uniform ionization states  enable us to make well-informed associations between $n_e$ and line emissivity. So that we are able to rely on these results,  we model and discuss the implications of density sampling relative to the critical densities in the local heating regime ($t > 200$ days) for the remainder of this work.

\subsection{Electron Critical Densities}\label{subsec:32}

Before we discuss in detail the relationship between electron density and emissivity, we first explain how to relate emissivity, electron density, and mass density within a shell of ejecta.  Given the near constant ionization fractions for Fe-group and IMEs, we assume that $n_{e} \propto \rho$, with an average of 2 free electrons per atom. We tested this assumption relative to the $n_e$  in \cite{blondin2023nebular} for the double detonation model against our spherically averaged density profile with the assumed electron density-mass density relation and found strong agreement, especially for the relevant velocity regime $<10,000 \mathrm{km/s}$. Given that the electron density is proportional to the mass density,  we directly convert between our simulated mass density for a region of the ejecta to electron density. 

In a low density environment we balance collisional excitation with collisional de-excitation and spontaneous emission to estimate the occupation $n_u$ of the upper state in an atom:
\begin{equation}
    n_l n_e q_{lu} = n_u (A_{ul}+ n_eq_{ul}),
\end{equation}
where $A_{ul}$ is the radiative transition probability from the upper to lower state and  $q_{lu}$ and $q_{ul}$ are the collisional excitation and de-excitation coefficients respectively. Emission line flux, the product of the number of spontaneous transitions occurring per unit time and the energy of the photon, is given by $ j_{lu} = E_{lu} A_{ul} n_u$, so substituting in the populations equation we find: 
\begin{equation}
 j_{lu} = \frac{n_l n_e q_{lu}}{A_{ul} + n_e q_{ul}} A_{ul} h \nu .
\end{equation}
If we characterize a specific number density (called the critical density) defined as 
 \begin{equation}
     n_{crit} = \frac{A_{ul}}{q_{ul}},
 \end{equation} 
 we can rewrite our emissivity as: 
\begin{equation}
    j_{lu}  \propto  \frac{n_l n_e}{1+ n_e/n_{crit}}.
\end{equation}For low densities within this regime, $n_e \ll n_{crit}$, $j _{lu} \approx n_l n_e$. If we take that the mass density is proportional to both $n_l$ and  $n_e$, $j _{lu} \approx \rho^2$. Now in the higher density limit ($n_e \gg n_{crit}$),  $j_{lu} \approx n_l n_{crit}$, so when $n_l \approx \rho$, $j_{lu} \approx \rho$.

Thus, above the critical electron density for a transition, the emission line flux scales with $\rho$, while below the critical density, the flux scales with $\rho ^2$. This will result in a change of the relationship of emissivity to mass density and, as we show, the shape and width of forbidden emission lines. 

\subsection{Modeling Forbidden Line Shapes}\label{subsec:33}

 Following \cite{Jerkstrand_2017}, we developed a code to model nebular line shapes by sampling ejecta mass density over regions within the ejecta corresponding to compositions of different element groups. As discussed by \cite{Jerkstrand_2017}, we treat the ejecta as a homologous expanding sphere with a maximum velocity $V_{max}$. First, we define $\Delta \nu = \nu - \nu_{0}$, with $\nu_{0}$ as the rest frequency of the emission line, and $\nu$ as the observed frequency. Each observed frequency $\nu$ corresponds to emission from a specific planar slice of the ejecta perpendicular to the observer's line of sight. If we take the x-direction as the observer's line of sight, this plane (in y-z space) is located at a projected velocity $V_x/c = (\Delta \nu/\nu_{0})$, or equivalently at a spatial position $x = V_{x}t$, where $t$ is the time since explosion. The thickness of this emitting slice along the line of sight $\Delta V_{x} = \Delta x/t$  is set by the intrinsic line width $\Delta V_0$ which is $\Delta V_0 \ll V_{max}$ so that each observed frequency effectively provides a two-dimensional cross-sectional integration (or ``scan") through the ejecta, sampling emission at that particular velocity slice, yielding a flux 
 \begin{equation}
    F_\nu \propto \int_{V(\nu)}^{V_{max}} j_0(V)VdV \propto \sum_j j\big(\rho(x_{ij})\big)  dA,
    \label{eq:flux}
\end{equation}
where $j_0$ is the co-moving emissivity.

  We sample over a spherical region larger than the maximum radius of the ejecta at a given time and integrate the density over planes perpendicular to the line of sight. Adding up the densities across all of the planes yields a dimensionless proxy for the flux, preserving the effects of the line shape due to the wake's asymmetries.  Following the construction for ejecta emission lines in \cite{Jerkstrand_2017}, we calculate the dimensionless flux quantity for an emissivity dependent solely on density as given in the second part of equation \ref{eq:flux}.

We divide the full velocity range of the ejecta into $i$ segments $v_i$ and multiply the density by the area vector which defines the plane. To account for all possible regimes relative to a line's critical density, we include options to sample the mass density linearly, quadratically or multiplied explicitly by the dimensionless emissivity:
 \begin{equation}
    j_{lu}  \approx  \frac{\rho n_e}{1+ n_e/n_{crit}},
    \label{eq:fullemiss}
\end{equation}
that incorporates $n_{crit}$. Across the plane we calculate:
\begin{equation}
    j_{ij} = \rho(x_{ij}) dA,
\end{equation}
substituting $\rho(x_{ij})^2$ for the squared regime and the full form of equation \ref{eq:fullemiss} for lines with critical densities close to the electron density at a given time. 

We sum  all of the points within the plane to get an integrated line emission value for the range of velocities corresponding with the bin $v_i$:
\begin{equation}
    J_i = \sum_j j_{ij}.
\end{equation}
Each of these bins correspond to one point of the overall emission line profile, broken into as many points as velocity bins. For our purposes, we use 100 velocity bins and sampling planes of size $100 \times 100$ to achieve a reasonable velocity resolution while keeping computation time low (less than 20 seconds to generate a single spectrum). 

This code calculates line shapes from any arbitrary 3D density field. We validated the code on densities corresponding to ejecta shapes with known analytical forms like the Gaussian and thin shell profiles derived in \cite{Jerkstrand_2017}. Our code agreed with the analytical solutions for the line shape, which we expected given the comparison between flux and our dimensionless quantity. We also tested the code on the analytical Gaussian density-velocity profile introduced in \cite{wong2024shocking} as well as expanded versions of the 1D model used in \cite{blondin2023nebular}. All of these validation tests resulted in correct line shapes. 

\subsection{Comparison to 2021aefx with Wake Free Ejecta}\label{subsec:34}

One result of our work is that when there is a constant mean ionization fraction, we can simply model the line shape and incorporate the transition from $j \propto \rho$ to $j \propto \rho^2$ as the ejecta electron density crosses $n_{crit}$ for various lines. 
\cite{gerardy2007signatures} established that the [Co III] $11.89 \mu m$ line is optically thin for  $5,000-10,000$K at $ > 200$ days after the explosion. Using values for the Einstein A coefficient and collision strengths at 6000 K from \cite{storey2016collision}, we find a critical density of $n_{crit} \approx 3.0 \times 10 ^5 \mathrm{cm}^{-3}$. Comparing with the calculated values of the electron density at 270 days post explosion from both \cite{blondin2023nebular} and our own assumption of 2 free electrons per atom in the interior finds electron densities in this range. Therefore, we expect the density sampling of the full expression for emissivity in equation \ref{eq:fullemiss} will  be  appropriate.

\begin{figure*}[ht]           
  \centering
  \includegraphics[width=\textwidth]{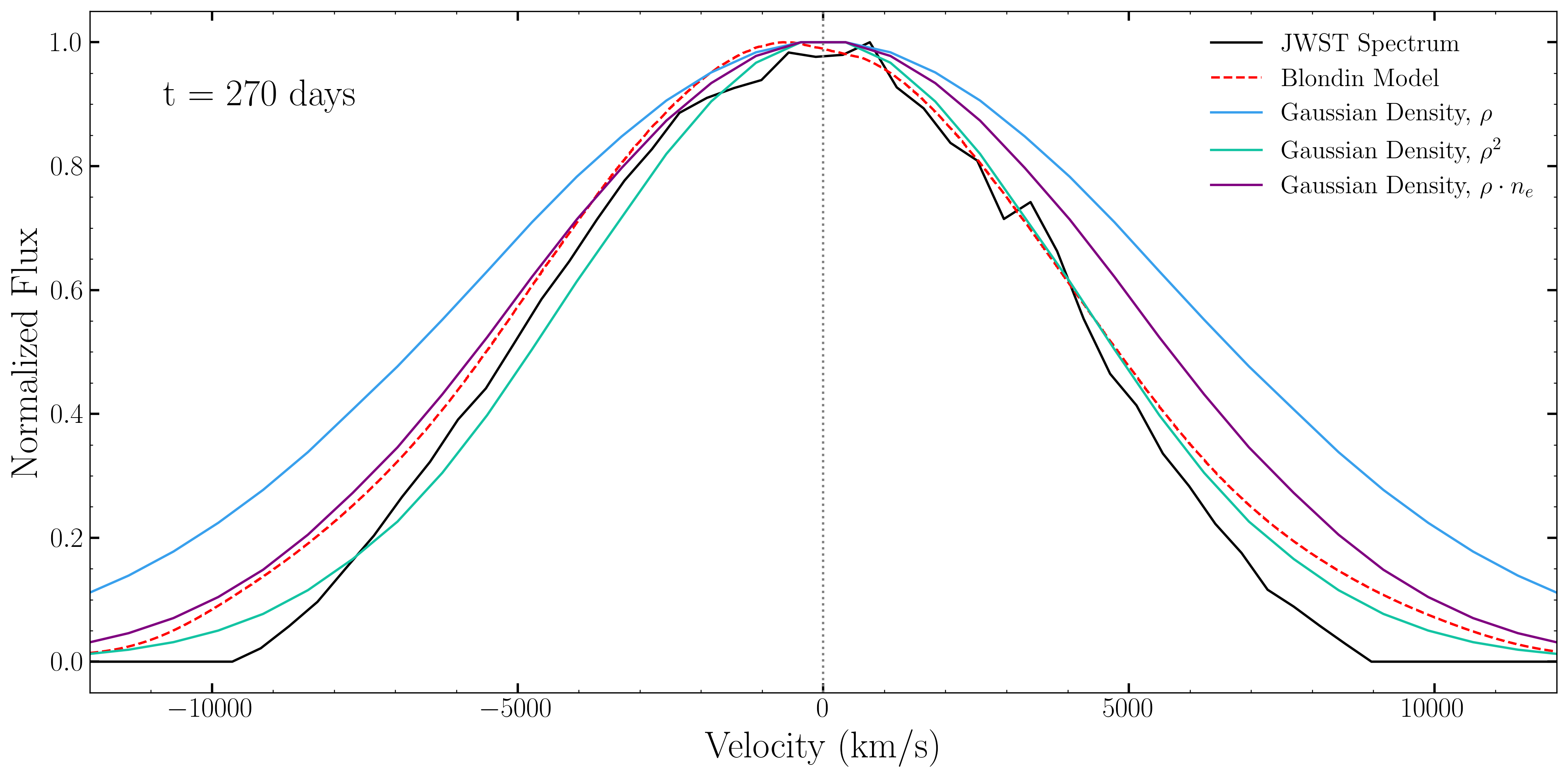}
  \caption{Comparison of the [Co III] $11.89 \mu m$ line shape from \cite{derkacy2023jwst}, Blondin 1D nLTE radiative transfer code output \citep{blondin2023nebular}, and our 3 sampling methods with an input of a spherically symmetric Gaussian density profile with equivalent ejecta mass and energy values as the model in \cite{blondin2023nebular} and \cite{gronow2021double}.
  \label{fig:compare}}
\end{figure*}

We test our assumptions by comparing to the line shape in recent JWST data \citep{kwok2023jwst,derkacy2023jwst} of SN 2021aefx. We start by calculating the line profiles with the Gaussian ejecta model from \cite{wong2024shocking} with the explosion parameters from \cite{gronow2021double}. We first compared the Gaussian ejecta model with the 1D model from \cite{gronow2021double} used in \cite{blondin2023nebular} as the primary `DBLEDT' model and found very good agreement.  We examine the [Co III] $11.89 \mu\mathrm{m}$ line at both 270 and 415 days post-explosion and compare all three of our density sampling methods for our spectral model to both the data and the 1D non-LTE spectrum in \cite{blondin2023nebular}. The comparison of JWST data originally from \cite{derkacy2023jwst}, the 1D radiative transfer code in \cite{blondin2023nebular}, and our three sampling methods using the Gaussian ejecta model from \cite{wong2024shocking} at 270 days is shown in Figure \ref{fig:compare}. We find good agreement between our work and \cite{blondin2023nebular} in particular for our $n_e$ sampled and $\rho ^2$ sampled profiles. It is clear that the data lies between our $\rho$ and $\rho ^2$ models as we predict and is well captured by our $n_e$ sampled model using the $n_{crit}$ appropriate for [Co III] $11.89 \mu\mathrm{m}$.

We also find, contrary to the predictions of \cite{gerardy2007signatures} and \cite{ashall2024jwst}, that the maximum line intensity does not follow the expected exponential decay of $^{56}\mathrm{Co}$, nor does it fall off like the twice the exponential decay (as we might predict if the line were truly sampling the density squared), but a value in between the two. We calculate a peak flux ratio of $\approx 7.3$ for continuum subtracted JWST spectra of the [Co III] $11.89 \mu\mathrm{m}$ line between 270 and 415 days post-explosion (contrary to the factor of 4 given in \cite{ashall2024jwst} from 255-415 days). From our unitless density sampled spectra, we find a flux ratio of $1.8$ and an expected decay factor of $\approx 3.7$ for the line from 270-415 days, which results in an overall decay factor of $\approx 6.6$, which is close to the raw peak ratio we found. We leave further study and validation of the exact flux ratio to future work.

Another outcome of this work came is verifying that 
\cite{wong2024shocking}'s Gaussian ejecta model works well in a variety of circumstances once the mass and explosion energies are constrained.   It closely follows our simulated ejecta density (excluding the companion interaction), the  D6 model of \cite{gronow2021double}, and used in conjunction with our spectral model code allows for quick estimates of line shapes for [Co III] $11.89 \mu\mathrm{m}$.

Our code in conjunction with the Gaussian ejecta model can be used in the future as a quick check of approximate spherically symmetric line shapes given the basic explosion parameters and the appropriate choice of $n_{crit}$. 

\section{Nebular Line Profiles with Wakes}\label{sec:wakes}

Satisfied by the validity of our approach, we now calculate line profiles using the perturbed density structures from the wake simulation. We span the full sphere by extrapolating our empirical density-radius relation at our largest simulated angle of  $\approx 80^\circ$ (where the ejecta is undisturbed by the wake) around the rest of the sphere. We then calculate   line profiles for various viewing angles, where the normalizations of density and flux differ in time as the ejecta expands.  The only time-changing consideration for the line shapes is the approach of the free electron densities to the critical densities of specific lines we investigate. Ionization structure and temperature as well as the balance and mechanism of  heating and cooling only mildly evolve over the nebular phase as discussed earlier in Section \ref{sec:emission}. 

\begin{figure*}[ht]           
  \centering
  \includegraphics[width=\textwidth]{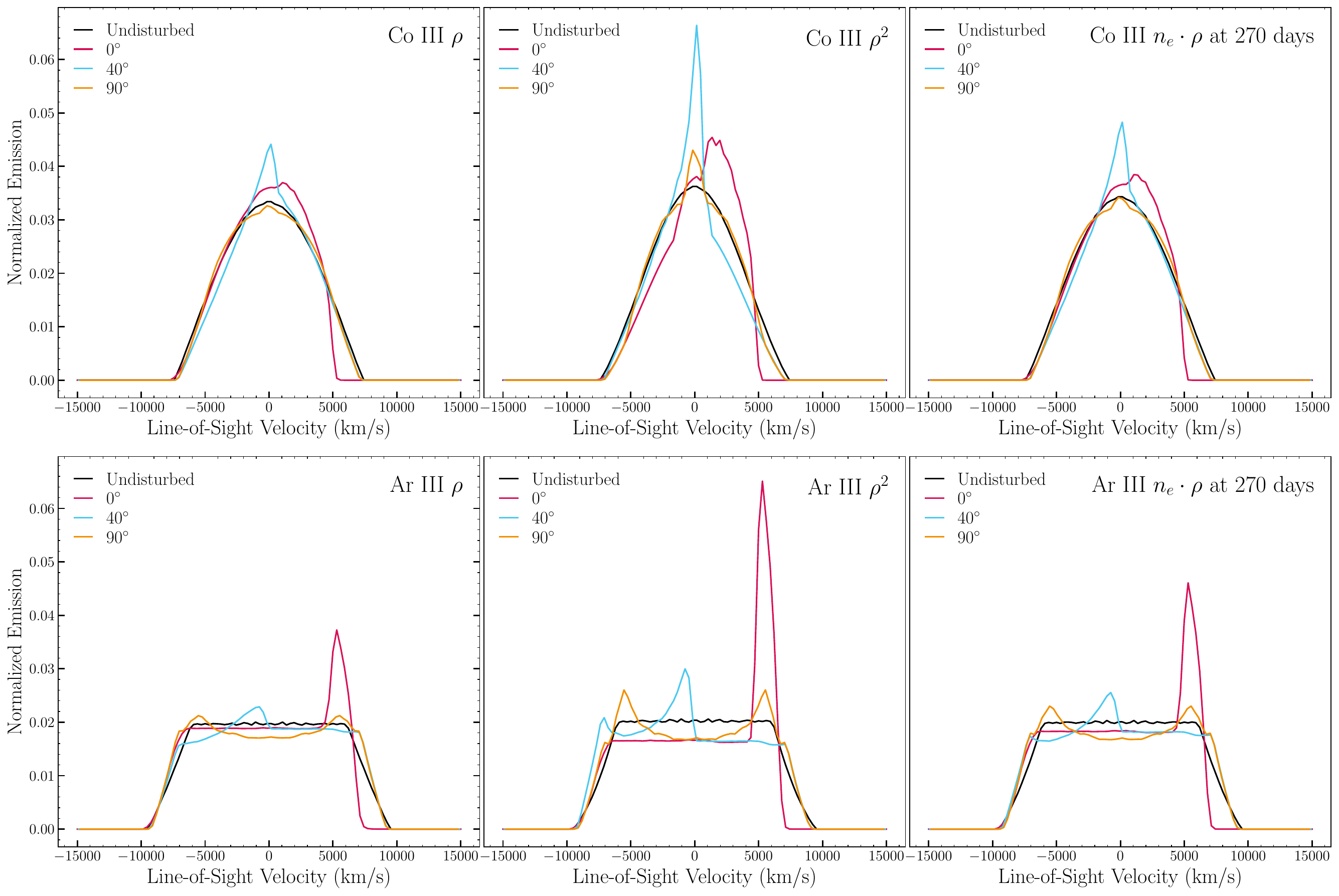}
  \caption{Modeled emission line profiles for [Co III] $11.89 \mu m$ (or the innermost velocity tracer as displayed in the top of Figure 2) on the top row and [Ar III] $8.99 \textit{$\mu$m}$ (the first shell-like tracer shown in the bottom of Figure 2). The solid black line shows the undisturbed (no-wake) ejecta modeled by the Gaussian fit of \cite{wong2024shocking}. In all panels the red line is the profile along the axis of the wake, the blue line is $40^\circ$ from the wake axis along the bow shock, and the orange line is $90^\circ$ from the wake axis. The first column uses linear sampling $j \propto \rho$. The second column uses squared sampling $j  \propto \rho^2$\, and the third column uses electron density sampling using the values of $n_{crit}$ for each ion at 270 days after explosion. \label{fig:spectra}}
\end{figure*}

Using our passive scalars, which serve as approximate composition tracers within our simulation, we isolate sections of our ejecta in regions dominated by particular element groups. We then calculate specific line shapes given our understanding of the velocity distributions of various elements within the wake-impacted ejecta. We focus on the two innermost sections of our ejecta, the region of material  that entered the simulation at $v \leq 7,000 \mathrm{km/s}$, and a shell directly adjacent to this region of material initially from $7,000 < v \leq 9,000 \mathrm{km/s}$. The mass densities multiplied by the values of each of these scalars were shown in Figure \ref{fig:tracer_density}. The low density wake and accumulation of material along the bow shock are characteristic asymmetries that will modify the emission line shapes. 

\subsection{\texorpdfstring{Co [III] 11.89 \textit{$\mu$m} Line Profiles}{Co [III] 11.89 micron Line Profiles}}\label{subsec:Co}

Prior work (\cite{gerardy2007signatures}, \cite{derkacy2023jwst}, \cite{kwok2024ground} ) highlighted [Co III] $11.89 \mu m$ as a strong, un-blended line that both traces the innermost ejecta and often displays asymmetries. It also traces the initial ${}^{56}\mathrm{Ni}$ density distribution, which is mostly confined to the innermost ejecta, and we associate with our $v \leq 7,000 \mathrm{km/s}$ tracer.

The top 3 panels of Figure \ref{fig:spectra} show [Co III] $11.89 \mu m$ line shapes (assuming a constant composition within our scalar-defined region of the ejecta) in the high, mid (near 270 days)  and low electron density regimes at 3 observer angles: along the explosion axis $\theta = 0$, perpendicular to that axis, and at an angle of $ \theta \approx 40^\circ$ (the opening angle of the bow shock from the ejecta moving around the companion). We also show the undisturbed (wake-free) line profile. 

Along the explosion axis, we notice a characteristic lack of high velocity material moving towards the observer, indicating the presence of the low density wake. When viewed along the bow shock, there remains material missing towards the line of sight and a build-up of material within the innermost region of the projected velocities, which is characterized by the overdensity along the bow shock. Perpendicular to the companion axis, we see a shape comparable to the undisturbed material as that makes up the bulk of the projected velocities. However, we notice a characteristic bump in the very center of the line as well as a symmetric switch between a slight under-prediction of material to over-prediction relative to the fully undisturbed density for larger angles. All three of the sampling schemes follow similar patterns, with the line narrowing and the asymmetries becoming more extreme from the linear to `electron density' ($n_e \cdot \rho$) sampling schemes and then again from the `electron density' to the squared sampling.

\subsection{\texorpdfstring{Shell-Like Emission (Ar [III] 8.99 \textit{$\mu$m})}{Shell-Like Emission (Ar [III] 8.99 micron)}}\label{subsec:Ar}

 We chose the relatively un-blended line [Ar III] $8.99 \textit{$\mu$m}$ as a representative of IME group emission that would be confined to a shell.  \cite{Jerkstrand_2017} showed that the thin shell approximation for an emitting region gives rise to a flat-topped line  profile.  [Ar III] $8.99 \textit{$\mu$m}$ has a characteristic flat-topped profile, well-documented asymmetries, and is a reasonable tracer of inner shell-like regions within ejecta \citep{ashall2024jwst,derkacy2023jwst,kwok2024ground}. \cite{gerardy2007signatures} discusses the physics of this line and its low optical depth throughout the nebular phase. Various supernovae are known to display tilted flat-top profiles for this line as well as other asymmetries like the clear double-peaked profile in 2005df \citep{gerardy2007signatures}. 

 From values given in \cite{yan2024fine}, we calculate $n_{crit} \approx  5 \times 10 ^5 \mathrm{cm}^{-3}$ for [Ar III] $8.99 \textit{$\mu$m}$. Like in the previous section with the [Co III] line, this critical density is comparable to the values of $n_e$ at late times \citep{blondin2023nebular}. We therefore expect the mid-regime electron density sampling to be the most accurate to the true line shape at 270 days post-explosion. 

As with the [Co III] line  profile, we show all three sampling schemes each at three different angles for the shell emission line in Figure \ref{fig:spectra}. Along the axis of the companion, we see a lack of high velocity material moving towards the observer, corresponding to the gap in the shell caused by the low-density wake. We then see a large spike in the profile, creating a single-horned shape when the ejecta density jumps up due to the material within the shell. The rest of the profile follows the same shape as the flat-topped profile in undisturbed ejecta, plotted alongside the profiles for each angle for convenience. For larger viewing angles,  a secondary bump appears corresponding to the over-abundance on the other side of the shell. The bump moves along the width of the line shape as a function of angle until reaching a full, symmetric double-horned shape along the line of sight perpendicular to the companion axis. Perhaps even more than the [Co III] line, the wake signatures on shell emission lines is striking. A comparison of these horn-like features to data should serve as a strong indicator of the presence of a wake. 

\subsection{First Velocity Moment}\label{subsec:velocity}

The first velocity moment (or intensity-weighted mean velocity) is a common diagnostic for measuring the bulk motion of emitting material along the line of sight. It is computed by integrating the observed flux-weighted velocity over a spectral line profile:
\begin{equation}
\langle v \rangle = \frac{\int v F(v)\, dv}{\int F(v)\, dv}.
\end{equation}
Physically, this moment captures the motion of the emitting ions weighted by their emissivity. Since expanding supernova ejecta produce Doppler-shifted spectral lines, asymmetries in the explosion geometry or in the distribution of radioactive elements (like ${}^{56} \mathrm{Ni}$) can result in net blueshifts or redshifts of the line centroid.  These shifts are used to infer whether more emitting material is moving toward or away from the observer, which in turn provides insight into explosion asymmetry, viewing angle, orbital velocities, and the momentum distribution of the ejecta. 

The first moment is related to physical momentum  when in the linear  density sampling scheme ($j \propto \rho $). It is similar to the total momentum divided by the total mass, and assuming a linear relationship with density, the first moment can serve as a proxy for the directional bias in momentum transport. 

\begin{figure}[ht]           
  \centering
  \includegraphics[width=0.45\textwidth]{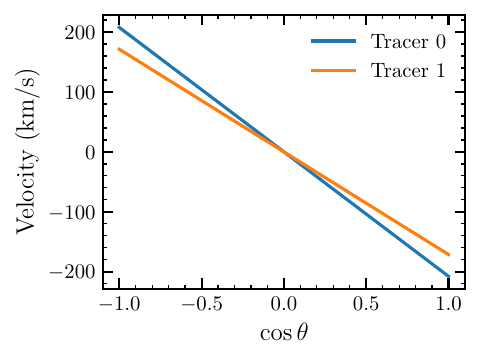}
  \caption{ The intensity-weighted velocity as a function of viewing angle. The blue line corresponds with the lowest velocity tracer and the orange line corresponds to the first shell like tracer.  \label{fig:moment}}
\end{figure}

Figure \ref{fig:moment}  shows the average intensity-weighted velocity (first velocity moment) for the two lowest velocity tracers as a function of viewing angle, parameterized by $cos\theta$ where $\theta$ is the angle between the observer’s line of sight and the explosion axis. The linear trend indicates a strong dipolar asymmetry, and the negative slope shows that asymmetries in small angles correspond to negative velocity shifts towards the observer, while the larger angles (around the back of the companion), ejecta is shifted away from the observer. The slope of the line can be seen as the projection of the ejecta momentum onto the line of sight. 
We also modeled the first velocity moment for our other sampling methods beyond just linear density sampling, where the physical analogy to momentum is lost. For the $\rho^2$ density we found that the sign of the slope flipped and the range of the velocity shift extended out to $\pm 600 \mathrm{km/s}$ . 

While these shifts are significant, they fall short of explaining the full range of velocity offsets observed in some Type Ia supernovae, which can exceed $1000 \mathrm{km/s}$ depending on line species and epoch (e.g., \cite{tucker2025merging}). This suggests that additional effects—such as large-scale asymmetries in the ${}^{56}\mathrm{Ni}$ distribution, off-center ignition, orbital velocities, or viewing-angle-dependent ionization—may be required to explain the observed velocities.

\section{Conclusions}\label{sec:conclusion}

This work demonstrates that companion interactions have a significant impact on late-time ($>200$ days) nebular phase line shapes  both for central and shell-like elemental distributions. As nearly all Type Ia supernova progenitor scenarios have a Roche-lobe filling companion at the time of the first explosion, our exposition of the way to detect the $10\%$ of ejecta that is modified should prove valuable.  Wakes also persist in the outer-most fast moving ejecta in the recently explored 
quadruple detonations \citep{boos2024type}. Though we have not explored wider binary configuration that are not Roche-lobe filling, the simplest expectation is a smaller effect due to less material being modified by the companion interaction. 


We present an angle-dependent study of the impact of companion interactions for both a shell-like distribution of material (typical of an IME like [Ar III]), and a deeper distribution closer to the core of the ejecta (typical of Fe-group elements like [Co III]). We find significant angle-dependent asymmetries in nebular phase line profiles, including strong peaks, horns, and double-horned like profiles that are distinctive enough to be identified in JWST observations of SNe Ia. Our work may provide an alternative or complement to some of the other explanations of line asymmetries present in recent JWST spectroscopy \citep{kwok2024ground, derkacy2023jwst, ashall2024jwst} as well as theoretical work \citep{blondin2023nebular,hoeflich2021measuring}. 

We also highlighted in Section \ref{sec:emission} that consideration of the electron critical density, $n_{crit}$, for a forbidden line motivates a better sampling approach, revealing that their shape changes depends on the proximity of the free electron density to  $n_{crit}$.
We predict that such a change in line shape should be detectable in current nebular phase observations assuming data are taken from substantially before and after the ejecta density crosses the threshold of $n_{crit}$. 
Using the convenient Gaussian ejecta model from \cite{wong2024shocking}, 
we show the expected line shape changes and compared them to the observed line shapes for 2021aefx in Section \ref{subsec:34}. Though this physics is in the full non-LTE radiative transfer codes, it has not been explicitly noted as a valuable reason for late time monitoring at a few epochs.

\begin{acknowledgments}
Acknowledgements: We thank Joe Hennawi, Gabriel Kumar, Lindsey Kwok, Chris Ni, Michael Tucker and Sunny Wong for beneficial conversations. 
This research benefited from interactions with a variety of researchers that were funded by the Gordon and Betty Moore Foundation through Grant GBMF5076. LJP is supported by a grant from the Simons Foundation (216179, LB) as well as a grant from the NASA Astrophysics Theory Program (ATP-80NSSC22K0725).
Use was made of computational facilities purchased with funds from the National Science Foundation (CNS-1725797) and administered by the Center for Scientific Computing (CSC). The CSC is supported by the California NanoSystems Institute and the Materials Research Science and Engineering Center (MRSEC; NSF DMR 2308708) at UC Santa Barbara.
This material is based upon work by KS supported by the U.S. Department of
Energy, Office of Science, Office of Advanced Scientific Computing Research, Department of
Energy Computational Science Graduate Fellowship under Award Number (DE-SC0024386). This report was prepared as an account of work sponsored by an agency of the
United States Government. Neither the United States Government nor any agency thereof, nor
any of their employees, makes any warranty, express or implied, or assumes any legal liability
or responsibility for the accuracy, completeness, or usefulness of any information, apparatus,
product, or process disclosed, or represents that its use would not infringe privately owned
rights. Reference herein to any specific commercial product, process, or service by trade name,
trademark, manufacturer, or otherwise does not necessarily constitute or imply its
endorsement, recommendation, or favoring by the United States Government or any agency
thereof. The views and opinions of authors expressed herein do not necessarily state or reflect
those of the United States Government or any agency thereof.
\end{acknowledgments}

%

\facilities{ UCSB Center for Scientific Computing at the California NanoSystems Institute (CNSI), Kavli Institute for Theoretical Physics}


\software{Astropy \citep{astropy2013, astropy2018}, 
          NumPy \citep{numpy}, 
          SciPy \citep{scipy}, 
          Athena++ \citep{athena++}}


\newpage
\appendix{}

\section{Low Velocity Ambiguities}\label{sec:A1}

Our simulation does not model the donor, the effects of stripping, or its gravity. Therefore, very low velocity (lower than the donor's escape velocity) ejecta is not well modeled. The ideal way to mitigate this would be to create a simulation similar to \cite{wong2024shocking} wherein the  authors explicitly simulated interactions between stripped material and low velocity ejecta, but run the scenario out to a much larger time and radius so that the inner ejecta and stripped material reaches homology. We leave this follow-up to future work and instead demonstrate a few physically-motivated approximations. 

\begin{figure}[ht]           
  \centering
  \includegraphics[width= 0.9\textwidth]{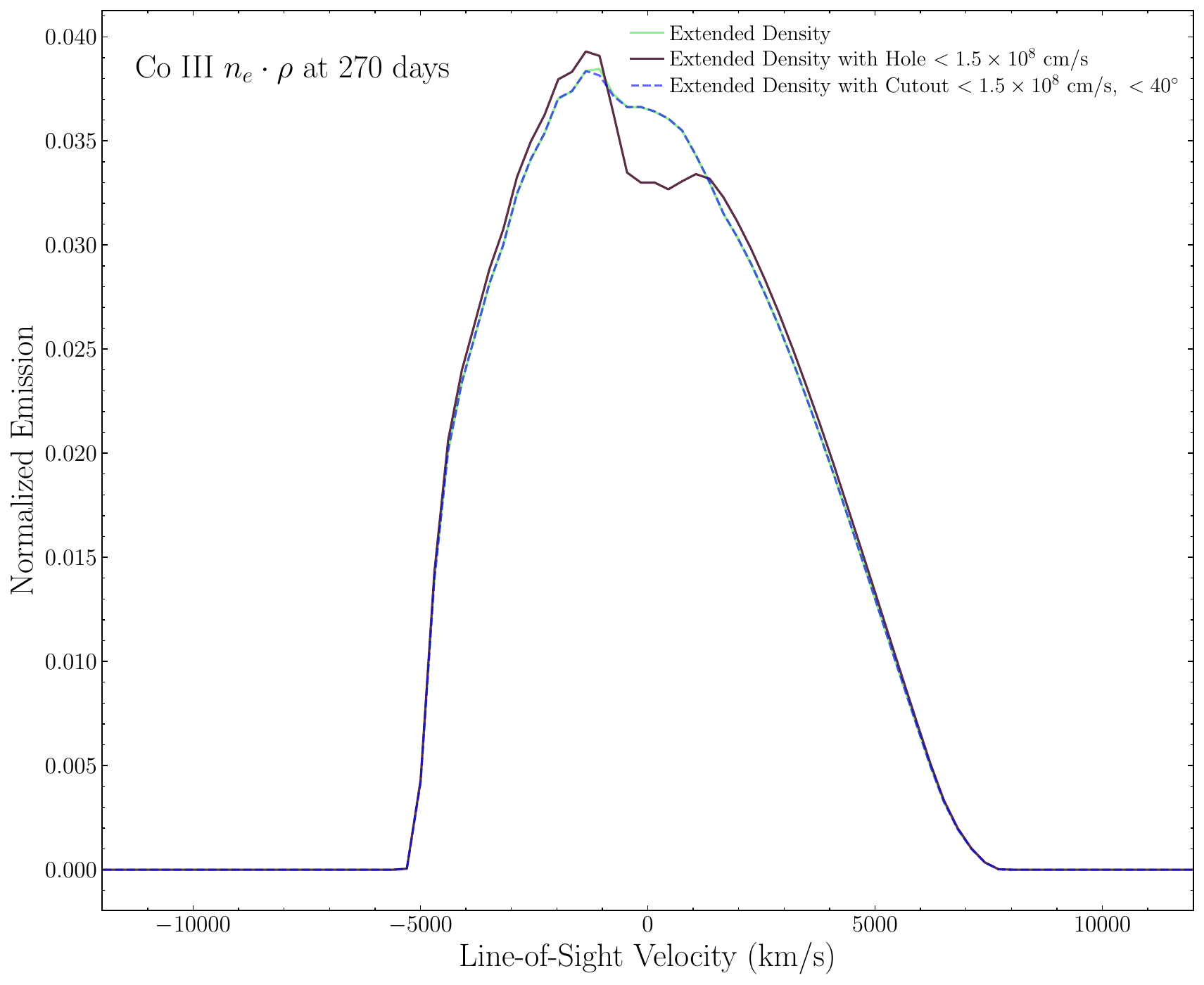}
  \caption{ Demonstrating the impact of 3 choices of innermost velocity scheme for electron density sampled [Co III] line. The gray line shows the most extreme case where all material at $\leq 1,500 \mathrm{km/s}$ is replaced with a hole. The blue line shows the case where a hole is cutout only within the angular extent of the bow shock ($\approx 40^\circ$). The green line shows an extension of our velocities at exactly $1,500 \mathrm{km/s}$ extrapolated inward to the edge of the simulation domain. \label{fig:lowv}}
\end{figure}

There are three feasible options to extrapolate our innermost ejecta at $\leq 1,500 \mathrm{km/s}$: following the Gaussian density fit from \cite{wong2024shocking}, extrapolating our velocity material at $\leq 1,500 \mathrm{km/s}$ to the innermost regions, and simply putting a hole or cutting out all material at $\leq 1,500 \mathrm{km/s}$ (either confined to a small angular region or not). The Gaussian density model does not capture any more detailed modeling of the donor interaction than our own ejecta density so we decided against this. We demonstrate the other two options in Figure \ref{fig:lowv}, showing that extending the density with a physically-motivated cutout corresponding to the solid angle taken up by stripped donor material is nearly indistinguishable to the extrapolated density without such a cutout at the resolution of our emission line models.  We chose to show this with the electron density sampling regime because the effect is more exaggerated in the squared regime and less exaggerated in the linear regime. The ``hole" and  in particular the cutout models would be appropriate given a strong presence of stripped He from the donor as suggested by \cite{wong2024shocking}. However, since we don't have concrete evidence of the extent to which this hole persists into the homologous expansion, we leave it to future work to investigate the full impact of these innermost velocities on nebular phase spectra.

\bibliography{sample631}{}

\begin{thebibliography}{}
\expandafter\ifx\csname natexlab\endcsname\relax\def\natexlab#1{#1}\fi
\providecommand{\url}[1]{\href{#1}{#1}}
\providecommand{\dodoi}[1]{doi:~\href{http://doi.org/#1}{\nolinkurl{#1}}}
\providecommand{\doeprint}[1]{\href{http://ascl.net/#1}{\nolinkurl{http://ascl.net/#1}}}
\providecommand{\doarXiv}[1]{\href{https://arxiv.org/abs/#1}{\nolinkurl{https://arxiv.org/abs/#1}}}

\bibitem[{Ashall {et~al.}(2024)Ashall, Hoeflich, Baron, Shahbandeh, Derkacy, Medler, Shappee, Tucker, Fereidouni, Mera, {et~al.}}]{ashall2024jwst}
Ashall, C., Hoeflich, P., Baron, E., {et~al.} 2024, The Astrophysical Journal, 975, 203

\bibitem[{{Astropy Collaboration} {et~al.}(2013)}]{astropy2013}
{Astropy Collaboration}, {et~al.} 2013, AA, 558, A33

\bibitem[{{Astropy Collaboration} {et~al.}(2018)}]{astropy2018}
---. 2018, AJ, 156, 123

\bibitem[{Bauer {et~al.}(2019)Bauer, White, \& Bildsten}]{bauer2019remnants}
Bauer, E.~B., White, C.~J., \& Bildsten, L. 2019, The Astrophysical Journal, 887, 68

\bibitem[{Blondin {et~al.}(2023)Blondin, Dessart, Hillier, Ramsbottom, \& Storey}]{blondin2023nebular}
Blondin, S., Dessart, L., Hillier, D.~J., Ramsbottom, C.~A., \& Storey, P.~J. 2023, Astronomy \& Astrophysics, 678, A170

\bibitem[{Boos {et~al.}(2024)Boos, Townsley, \& Shen}]{boos2024type}
Boos, S.~J., Townsley, D.~M., \& Shen, K.~J. 2024, The Astrophysical Journal, 972, 200

\bibitem[{Boos {et~al.}(2021)Boos, Townsley, Shen, Caldwell, \& Miles}]{boos2021multidimensional}
Boos, S.~J., Townsley, D.~M., Shen, K.~J., Caldwell, S., \& Miles, B.~J. 2021, The Astrophysical Journal, 919, 126

\bibitem[{De {et~al.}(2019)De, Kasliwal, Polin, Nugent, Bildsten, Adams, Bellm, Blagorodnova, Burdge, Cannella, {et~al.}}]{de2019ztf}
De, K., Kasliwal, M.~M., Polin, A., {et~al.} 2019, The Astrophysical Journal Letters, 873, L18

\bibitem[{De {et~al.}(2020)De, Kasliwal, Tzanidakis, Fremling, Adams, Aloisi, Andreoni, Bagdasaryan, Bellm, Bildsten, {et~al.}}]{de2020zwicky}
De, K., Kasliwal, M.~M., Tzanidakis, A., {et~al.} 2020, The Astrophysical Journal, 905, 58

\bibitem[{DerKacy {et~al.}(2023)DerKacy, Ashall, Hoeflich, Baron, Shappee, Baade, Andrews, Bostroem, Brown, Burns, {et~al.}}]{derkacy2023jwst}
DerKacy, J., Ashall, C., Hoeflich, P., {et~al.} 2023, The Astrophysical Journal Letters, 945, L2

\bibitem[{Desai {et~al.}(2025)Desai, Haggerty, Shappee, Tucker, Davis, Ashall, Chomiuk, Gootkin, Caprioli, Bret, {et~al.}}]{desai2025plasma}
Desai, D.~D., Haggerty, C.~C., Shappee, B.~J., {et~al.} 2025, arXiv preprint arXiv:2504.15335

\bibitem[{El-Badry {et~al.}(2023)El-Badry, Shen, Chandra, Bauer, Fuller, Strader, Chomiuk, Naidu, Caiazzo, Rodriguez, {et~al.}}]{el2023fastest}
El-Badry, K., Shen, K.~J., Chandra, V., {et~al.} 2023, arXiv preprint arXiv:2306.03914

\bibitem[{Gerardy {et~al.}(2007)Gerardy, Meikle, Kotak, H{\"o}flich, Farrah, Filippenko, Foley, Lundqvist, Mattila, Pozzo, {et~al.}}]{gerardy2007signatures}
Gerardy, C.~L., Meikle, W., Kotak, R., {et~al.} 2007, The Astrophysical Journal, 661, 995

\bibitem[{Gronow {et~al.}(2021)Gronow, Collins, Sim, \& R{\"o}pke}]{gronow2021double}
Gronow, S., Collins, C.~E., Sim, S.~A., \& R{\"o}pke, F.~K. 2021, Astronomy \& Astrophysics, 649, A155

\bibitem[{Harris {et~al.}(2020)}]{numpy}
Harris, C.~R., {et~al.} 2020, Nature, 585, 357–362

\bibitem[{Hoeflich {et~al.}(2021)Hoeflich, Ashall, Bose, Baron, Stritzinger, Davis, Shahbandeh, Anand, Baade, Burns, {et~al.}}]{hoeflich2021measuring}
Hoeflich, P., Ashall, C., Bose, S., {et~al.} 2021, The Astrophysical Journal, 922, 186

\bibitem[{Jerkstrand(2017)}]{Jerkstrand_2017}
Jerkstrand, A. 2017, Spectra of Supernovae in the Nebular Phase (Springer International Publishing), 795–842, \dodoi{10.1007/978-3-319-21846-5_29}

\bibitem[{Kasen {et~al.}(2004)Kasen, Nugent, Thomas, \& Wang}]{kasen2004could}
Kasen, D., Nugent, P., Thomas, R., \& Wang, L. 2004, The Astrophysical Journal, 610, 876

\bibitem[{Kwok {et~al.}(2023)Kwok, Jha, Temim, Fox, Larison, Camacho-Neves, Brenner~Newman, Pierel, Foley, Andrews, {et~al.}}]{kwok2023jwst}
Kwok, L.~A., Jha, S.~W., Temim, T., {et~al.} 2023, The Astrophysical Journal Letters, 944, L3

\bibitem[{Kwok {et~al.}(2024)Kwok, Siebert, Johansson, Jha, Blondin, Dessart, Foley, Hillier, Larison, Pakmor, {et~al.}}]{kwok2024ground}
Kwok, L.~A., Siebert, M.~R., Johansson, J., {et~al.} 2024, The Astrophysical Journal, 966, 135

\bibitem[{Kwok {et~al.}(2025)Kwok, Singh, Jha, Blondin, Dastidar, Larison, Miller, Andrews, Andrews, Anupama, {et~al.}}]{kwok2025jwst}
Kwok, L.~A., Singh, M., Jha, S.~W., {et~al.} 2025, arXiv preprint arXiv:2505.02944

\bibitem[{Liu {et~al.}(2023)Liu, R{\"o}pke, \& Han}]{liu2023type}
Liu, Z.-W., R{\"o}pke, F.~K., \& Han, Z. 2023, Research in Astronomy and Astrophysics, 23, 082001

\bibitem[{Marietta {et~al.}(2000)Marietta, Burrows, \& Fryxell}]{marietta2000type}
Marietta, E., Burrows, A., \& Fryxell, B. 2000, The Astrophysical Journal Supplement Series, 128, 615

\bibitem[{Milne {et~al.}(1999)Milne, Leising, {et~al.}}]{milne1999positron}
Milne, P., Leising, M., {et~al.} 1999, The Astrophysical Journal Supplement Series, 124, 503

\bibitem[{Polin {et~al.}(2021)Polin, Nugent, \& Kasen}]{polin2021nebular}
Polin, A., Nugent, P., \& Kasen, D. 2021, The Astrophysical Journal, 906, 65

\bibitem[{Pollin {et~al.}(2024)Pollin, Sim, Pakmor, Callan, Collins, Shingles, R{\"o}pke, \& Srivastav}]{pollin2024fate}
Pollin, J., Sim, S., Pakmor, R., {et~al.} 2024, Monthly Notices of the Royal Astronomical Society, 533, 3036

\bibitem[{Prust {et~al.}(2025)Prust, Kumar, \& Bildsten}]{prust2025ejecta}
Prust, L.~J., Kumar, G., \& Bildsten, L. 2025, The Astrophysical Journal, 982, 60

\bibitem[{Shen {et~al.}(2021)Shen, Blondin, Kasen, Dessart, Townsley, Boos, \& Hillier}]{shen2021non}
Shen, K.~J., Blondin, S., Kasen, D., {et~al.} 2021, The Astrophysical Journal Letters, 909, L18

\bibitem[{Shen {et~al.}(2018)Shen, Boubert, G{\"a}nsicke, Jha, Andrews, Chomiuk, Foley, Fraser, Gromadzki, Guillochon, {et~al.}}]{shen2018three}
Shen, K.~J., Boubert, D., G{\"a}nsicke, B.~T., {et~al.} 2018, The Astrophysical Journal, 865, 15

\bibitem[{Sobolev(1957)}]{sobolev1957diffusion}
Sobolev, V. 1957, Soviet Astronomy, Vol. 1, p. 678, 1, 678

\bibitem[{Stone {et~al.}(2020{\natexlab{a}})Stone, Tomida, White, \& Felker}]{Stone2020}
Stone, J.~M., Tomida, K., White, C.~J., \& Felker, K.~G. 2020{\natexlab{a}}, The Astrophysical Journal Supplement Series, 249, 4, \dodoi{10.3847/1538-4365/ab929b}

\bibitem[{Stone {et~al.}(2020{\natexlab{b}})Stone, Tomida, White, \& Felker}]{athena++}
---. 2020{\natexlab{b}}, The Athena++ Adaptive Mesh Refinement Framework.
\newblock \doarXiv{2005.06651}

\bibitem[{Storey \& Sochi(2016)}]{storey2016collision}
Storey, P., \& Sochi, T. 2016, Monthly Notices of the Royal Astronomical Society, 459, 2558

\bibitem[{Tanikawa {et~al.}(2019)Tanikawa, Nomoto, Nakasato, \& Maeda}]{tanikawa2019double}
Tanikawa, A., Nomoto, K., Nakasato, N., \& Maeda, K. 2019, The Astrophysical Journal, 885, 103

\bibitem[{Tucker(2025)}]{tucker2025merging}
Tucker, M.~A. 2025, Monthly Notices of the Royal Astronomical Society: Letters, 538, L1

\bibitem[{Virtanen {et~al.}(2020)}]{scipy}
Virtanen, P., {et~al.} 2020, Nature Methods, 17, 261–272

\bibitem[{Wong \& Bildsten(2023)}]{wong2023dynamical}
Wong, T. L.~S., \& Bildsten, L. 2023, The Astrophysical Journal, 951, 28

\bibitem[{Wong {et~al.}(2024)Wong, White, \& Bildsten}]{wong2024shocking}
Wong, T. L.~S., White, C.~J., \& Bildsten, L. 2024, The Astrophysical Journal, 973, 65

\bibitem[{Yan \& Babb(2024)}]{yan2024fine}
Yan, P.-G., \& Babb, J.~F. 2024, The Astrophysical Journal, 961, 43

\end{thebibliography}
\bibliographystyle{aasjournal}



\end{document}